\documentstyle[11pt,aaspp4,flushrt]{article}
\newcommand{\NH}{{$N_{\rm H}$}}
\newcommand{\LX}{{$L_{\rm X}$}}

\newcommand{\eps}{ergs s$^{-1}$}
\newcommand{\pcm}{cm$^{-2}$}
\newcommand{\ps}{s$^{-1}$}

\newcommand{\asca}{{\it ASCA}}
\newcommand{\rosat}{{\it ROSAT}}

\newcommand{\gtsima}{$\; \buildrel > \over \sim \;$}
\newcommand{\simgt}{\lower.5ex\hbox{\gtsima}}
\newcommand{\ltsima}{$\; \buildrel < \over \sim \;$}
\newcommand{\simlt}{\lower.5ex\hbox{\ltsima}}

\slugcomment{To appear in {\it The Astrophysical Journal (Letters).}}

\lefthead{Terashima et al.}
\righthead{Iron K Variability in NGC 4579}

\begin{document}

\title{Iron K line Variability in the Low-Luminosity AGN NGC 4579}

\author{Yuichi Terashima\altaffilmark{1},
Luis C. Ho\altaffilmark{2}, 
Andrew F. Ptak\altaffilmark{3},
Tahir Yaqoob\altaffilmark{1,4},
Hideyo Kunieda\altaffilmark{5},
Kazutami Misaki\altaffilmark{5},
and Peter J. Serlemitsos\altaffilmark{1}
}

\altaffiltext{1}{NASA Goddard Space Flight Center, Code 662,
Greenbelt, MD 20771}

\altaffiltext{2}{The Observatories of the Carnegie Institution of
Washington, 813 Santa Barbara St., Pasadena, CA 91101-1292}

\altaffiltext{3}{Department of Physics, Carnegie Mellon University, 5000 Forbes
Ave., Pittsburgh, PA 15213}

\altaffiltext{4}{Universities Space Research Association}

\altaffiltext{5}{Institute of Space and Astronautical
Science, Yoshinodai 3-1-1, Sagamihara, Kanagawa 229-8510, Japan}

\begin{abstract}

We present results of new {\asca} observations of the low-luminosity
AGN (LLAGN) NGC 4579 obtained on 1998 December 18 and 28, and we
report on detection of variability of an iron K emission line. The
X-ray luminosities in the 2--10 keV band for the two observations are
nearly identical ({\LX} $\approx$ 2$\times10^{41}$ {\eps}), but they
are $\sim$35\% larger than that measured in 1995 July by Terashima et
al. An Fe K emission line is detected at $6.39\pm0.09$ keV
(source rest frame) which is lower than the line energy
$6.73^{+0.13}_{-0.12}$ keV in the 1995 observation. If we fit the Fe
lines with a blend of two Gaussians centered at 6.39 keV and 6.73 keV,
the intensity of the 6.7 keV line decreases, while the intensity of
the 6.4 keV line increases, within an interval of 3.5 yr. This
variability rules out thermal plasmas in the host galaxy as the origin
of the ionized Fe line in this LLAGN.  The detection and
variability of the 6.4 keV line indicates that cold matter subtends a
large solid angle viewed from the nucleus and that it is located
within $\sim1$ pc from the nucleus.  It could be identified with an
optically thick standard accretion disk.  If this is the case, a
standard accretion disk is present at the Eddington ratio of $L_{\rm
Bol}$/$L_{\rm Eddington} \sim 2\times10^{-3}$. A broad
disk-line profile is not clearly seen and the structure of the
innermost part of accretion disk remains unclear.

\end{abstract}


\section{Introduction}

Fe line emission in AGNs is an X-ray spectral feature produced through
reprocessing by the matter surrounding the nucleus, and it can be used
as a probe of the matter in the vicinity of the central black hole,
such as accretion disks.  Broad Fe emission lines have been detected
in a number of Seyfert 1 galaxies, and they have been interpreted as
arising from a relativistic accretion disk (e.g. Tanaka et al. 1995;
Nandra et al. 1997). Fe K emission has also been detected in several
low-luminosity AGNs (LLAGNs).  The origin of Fe lines in these
objects, however, is unclear.  The line centroid energy of some LLAGNs
is close to 6.7 keV (Ishisaki et al. 1996; Serlemitsos et al. 1996;
Iyomoto et al. 1997; Terashima et al. 1998a, hereafter T98; Roberts,
Warwick, \& Ohashi 1999), which is higher than 6.4 keV as seen in
Seyfert 1s, while some objects show a line at 6.4 keV (Terashima,
Kunieda, \& Misaki 1999). Possible origins of the Fe line at 6.7 keV
include an ionized standard accretion disk, a thermal plasma in the
outer region of an advection-dominated accretion flow (Narayan, \&
Raymond 1999), an ionized absorber and emitter along the line-of-sight
(Pellegrini et al. 2000), or a thermal plasma in the host galaxy (Ptak
et al. 2000).  If the higher line energy is due to ionized iron, the
problem arises as to how to achieve a high degree of ionization given
the low luminosities and low accretion rates in LLAGNs. The
variability of Fe lines provides information on the location of the
line-emitting region, and thus constraints on the origin of the lines. 
Studies of Fe lines in LLAGNs are also of great importance to
understand the structure of accretion disks in low luminosity objects.

NGC 4579 contains an LLAGN with {\LX}(2--10 keV) =
$1.5\times10^{41}$~{\eps} (we assume a distance of 16.8 Mpc; Tully
1988). An {\asca} observation was made on 1995 June 25, and an Fe K
line is detected at $6.73^{+0.13}_{-0.12}$ keV (source rest frame) with an
equivalent width (EW) of $490^{+180}_{-190}$ eV and a width of
$0.17^{+0.11}_{-0.12}$ keV (T98).  We observed NGC 4579 again with
{\asca} in 1998 to search for possible variability of the Fe line. In
this {\it Letter}, we report the change of the line centroid energy
associated with a continuum luminosity increase of about 35\%.

\section{The {\it ASCA} Data}

  We observed NGC 4579 on 1998 Dec. 18 and 28 with the {\asca}
satellite (Tanaka, Inoue, \& Holt 1994). The two Solid-State Imaging
Spectrometers (SIS0 and SIS1; Burke et al. 1994) were operated in the
1CCD Faint mode. The two Gas Imaging Spectrometers (GIS2 and GIS3;
Ohashi et al. 1996; Makishima et al. 1996) were operated in the 
nominal pulse height mode.

X-ray spectra were accumulated from circular regions with radii of
4$^\prime$ for the SIS and 6$^\prime$ for the GIS. Background spectra
were taken from a source-free region in the same field. The effective
exposure times, after data screening, were 18.6 ks for each SIS and
19.6 ks for each GIS on December 18, and 17.6 ks for each SIS and 19.6
ks for each GIS on December 28. The count rates were similar for both
observations: 0.14 counts s$^{-1}$ for SIS and 0.13 counts s$^{-1}$
for GIS, after background subtraction. The spectra from SIS0 and SIS1
were combined, as were those from GIS2 and GIS3. We fitted the SIS0
plus SIS1 spectra and the GIS2 plus GIS3 spectra simultaneously. It is
known that SIS and GIS spectra at the low-energy band ($<$1 keV) are
inconsistent with each other due to accumulated radiation damage of
the SIS (see Yaqoob et al. 2000). In order to minimize the uncertainty
in the low-energy band, we used the region 1.0--10 keV for the SIS. We
confirmed that the SIS and GIS data within this band yield
statistically consistent spectral fits.

\section{Results}

% (1) averaged spectrum

We fitted the combined spectrum of the two observations in 1998.  At
first we fitted the spectrum with a power law plus Gaussian model and a
thermal bremsstrahlung plus Gaussian model.  The former model provided
an acceptable fit with $\chi^2=201.0$ for 227 degrees of freedom
(dof). The latter model resulted in a significantly worse fit with
$\chi^2=302.7$ for 227 dof. Spectral features of a soft thermal
plasma, detected in the 1995 observation, were not clearly
seen in the residuals. We tried to add a Raymond-Smith thermal plasma
component (hereinafter RS) to the power law plus Gaussian model. The
temperature and abundance of the RS component were fixed at the values
of T98 ($kT$=0.90 keV and abundance = 0.5 solar), and the normalization
was left to vary. The absorption column density for the RS component
was fixed at the Galactic value ({\NH} = $3.1\times10^{20}$ {\pcm};
Murphy et al. 1996). The $\chi^2$ improved only by $\Delta \chi^2$ =
1.0. Thus we found no clear evidence for the presence of soft thermal
emission in the 1998 observation. This is probably for the
following reasons. First, the flux of the hard component has brightened
by a factor of 1.35 from 1995, and thus the soft thermal component has been
diluted. Second, we ignored the soft energy band below 1 keV of the
SIS data in the fit, where the soft component is expected to be more
prominent. The spectral slope of the hard component is nearly 
unaffected by the addition of the soft thermal component, while the
absorption column density depends on it. We use the power law + Gaussian + RS 
model in the following analysis.

	\placefigure{fig-1}

	\placefigure{fig-2}

	\placetable{tbl-1}

The best-fit parameters for the continuum and X-ray fluxes in the
2--10 keV band are shown in Table 1, as well as the fitting results of
the 1995 data taken from T98. The Fe line parameters are summarized in
Table 2. The SIS and GIS spectra (along with the best-fit continuum
model), and ratios of data to the best-fit continuum around Fe lines
in the 1998 observation are shown in Figure 1. Ratios of data to the
best-fit continuum for the 1995 observation are also shown in Figure 1
for comparison.  The hard X-ray luminosity in the 2--10 keV band ({\LX}
= $2.0\times10^{41}$ {\eps}) is $\sim$35\% larger in 1998 than in 1995. 
Although the best-fit spectral slope has slightly steepened, the
statistical errors overlap each other.  No significant change of the
absorption column density is seen either, although the errors are
large because of the limited energy band of SIS used in the analysis.
An intriguing spectral change is the centroid energy of the the Fe
line. The line center energy varied from $6.73^{+0.13}_{-0.12}$ keV in
1995 to $6.39\pm0.09$ keV in 1998. Confidence contours for the line
center energy versus normalization for the two observations are shown
in Figure 2. The change of the line center energy is significant at
more than 90\% confidence level for two interesting parameters. Note
that the calibration uncertainty of the energy scale is less than 1\%. 
We examined the gain uncertainties by using the gold edge around 2.3
keV and the Cu-K line of the GIS detector background and found no
evidence for a significant gain shift larger than the 1\% level.  We
also confirmed that all the four detectors provided Fe K line center
energies consistent with each other within 1\% using a recent
observation of the bright AGN NGC 5548 in 1998 June and July.

%[$>$2.3\% gain shift is rejected by chi-square fit to the edge and
%Cu-K line.])

	\placetable{tbl-2}

	\placefigure{fig-3}

% double Gaussian fit

The change of the line center energy could be caused by a change of
the ionization state of the line emitter or by variability of multiple
emission lines with different line centroid energies. In order to
quantify the second possibility, we modeled the line feature with a
combination of two Gaussians. We fixed the line center energies and
widths of each line at the best-fit values in 1995 and 1998. The
obtained EWs and intensities are shown in Table 2 and Figure 3. The
intensity and EW of the 6.7 keV line decreased significantly. The
intensity of the 6.4 keV line significantly increased, while the EW
marginally increased.  The continuum parameters in this fit are same
as in the single Gaussian model fits shown in Table 1.  If we assume
zero width for the 6.7 keV line instead of the best-fit value in 1995,
the conclusion on line variability is unchanged.

% variability between two observations (10 days apart)

Finally, we fitted the spectra of the two observations in 1998
separately to search for spectral variability on a time scale of 10
days. We used the same model in the fit of the combined spectrum. The
Fe line was modeled with a single Gaussian. The obtained spectral
parameters are also summarized in Table 1 and Table 2. The X-ray
luminosity in the 2--10 keV band is slightly ($\sim$15\%) higher in
the first observation compared to the second.  We found no clear
spectral change between these two observations.

\section{Discussion}

We found that the center energy of the Fe K line in NGC 4579 decreased
in 3.5 yrs, while the luminosity in the 2--10 keV band increased by
35\%. We examined two models to represent the observed variability: a
single Gaussian model and a double Gaussian model.  In the double
Gaussian model, the center energies were fixed at 6.39 keV and 6.73
keV. The results of these model fits indicate that the line intensity
of the 6.7 keV line varied within an interval of 3.5 yr. This
variability indicates that the emitter of the Fe line at 6.7 keV is
located within $\sim1$ pc from the nucleus, and thus rules out hot gas
in the host galaxy (starburst activity and/or ridge emission, which is
diffuse X-ray emission with $kT\sim5-10$ keV, as observed in our
Galaxy) as the origin of the 6.7 keV Fe line. On the other hand, the
Fe line at 6.4 keV has an origin in fluorescence in cold or slightly
ionized matter illuminated by a central X-ray source (Makishima 1986). 
Thus, the observed Fe lines at 6.4 keV and 6.7 keV most likely
originate from the AGN itself rather than from hot gas further out in
the host galaxy. The relatively strong 6.4 keV line in 1998 (EW =
$250^{+105}_{-95}$ eV) indicates that cold matter subtends a large
solid angle viewed from the nucleus (e.g., Gerorge \& Fabian 1991). A
part of the line could come from an obscuring torus which is assumed
in unified models of AGNs (e.g., Antonucci 1993). If the torus is
present and located at further than 1 pc from the center, the Fe line
from it does not vary within an interval of $\sim$ 3 yr. Then, the
torus contribution can be at most $8.5\times10^{-6}$ photons$^{-1}$
{\ps} {\pcm}, which is the measured 90\% confidence upper limit on the
constant 6.4 keV line from the 1995 observation (see Table 2).  This
line flux corresponds to an EW of 155 eV for the continuum level in
1998.  The rest of the 6.4 keV line should come from the region closer
to the nucleus.  Such a Fe line emitter could be identified with an
optically thick accretion disk. If this is the case, a standard
accretion disk should exist even at the low Eddington ratio of NGC
4579 ($L_{\rm Bol}/L_{\rm Eddigton} = 2.0\times10^{-3}$; Ho 1999). The
presence of an optically thick accretion disk is consistent with the
estimate of the transition radius from an advection dominated
accretion flow to a standard disk ($\sim$100 Schwarzschild radii) by
Quataert et al. (1999). However, the suggestion of a truncated disk
might not be appropriate
 to achieve the observed large equivalent width ($\approx$250
eV). Therefore, cold matter beyond the accretion disk, such as an
obscuring torus, must contribute somewhat to the observed EW of the
6.4 keV line, if the truncated disk interpretation is correct.

The origin of the 6.7 keV line seen in 1995 still remains puzzling,
however. If the Fe lines in 1995 and 1998 originated from the same
region, the decrease of ionization state responding to the small flux
increase of only 35\% is difficult to be explained by photoionization
under an assumption of a constant density. If the ionized matter is
located far from the nucleus, the observed behavior of the Fe line
could be due to a time lag between the continuum and Fe line
variability.  The X-ray flux in the 0.5--2 keV band observed with the
{\rosat} PSPC in 1991 Dec. is a factor of 2.3 higher than that of the
first {\asca} observation in 1995 June (Ptak et al. 1999). Therefore,
the decrease of ionization state might be explained by a time lag
effect, if the Fe line emitter is located 1 pc away form the nucleus.
In order to explain the change of ionization state from \ion{Fe}{25}
(6.7 keV line) to \ion{Fe}{17} or less (6.4 keV line), the ionization
parameter $\xi = L/nR^2$ should change from $\log \xi \approx 3$ to 2
(e.g., Kallman, \& McCray 1982), where $L$, $n$, $R$ are the
luminosity of ionizing photons, the number density of photoionized
matter, and the distance from the ionizing source to the photoionized
matter, respectively. The observed variability, however, is not enough
to explain the change of the ionization parameter unless the density
and/or geometry of the ionized matter also changed, or large-amplitude
variability occurred during the unobserved span.  Long-term
variability of a factor 10 is reported in only a few objects (M81,
Pellegrini et al. 2000; and possibly M51, Terashima et al. 1998b).

Alternatively, the 6.4 keV and 6.7 keV lines may originate from
different regions.  If most of the 6.4 keV line comes from the torus
and the 6.7 keV line is from different photoionized matter, the
disappearance of the 6.7 keV line might be understood by increasing
the ionization state such that the gas is almost fully ionized. In
this case, the 6.4 keV line could vary in response to the continuum
variability with some time lag which depends on the size and geometry
of the torus.  However, the small flux change, again, is not enough to
change the ionization state so drastically. 

Thus, it is conceivable that the physical conditions (density and/or
geometry) of the ionized matter changed between the two observations
whether the 6.4 keV and 6.7 keV lines originate from a single region
or different regions. Such a change could be caused by a change of the
ionization balance in the surface layer of the standard accretion
disk, or in the boundary region between the standard disk and the
advection-dominated flow.  It is worth noting that there are several
Seyfert 1 galaxies showing the change of Fe K line centroid energy
accompanied by small variations in the continuum luminosity (Weaver,
Gelbord, \& Yaqoob 2000). The cause of such variability, however, is
yet to be understood.  We note also that a drastic change of Fe line
profile is observed during a flare in MCG--6--30--15 (Iwasawa et
al. 1999), althogh the time
scale they observed is much shorter than ours.  The peak energy of the
Fe line during the flare was around 5 keV and such a profile change can
be interpreted as being due to a flare occuring very close to the
black hole (Iwasawa et al. 1999).

The observed line does not show an obvious asymmetric profile skewed
toward lower energies, a signature expected from a relativistic disk
(e.g., Fabian et al. 1989) and one which is observed in many Seyfert 1
nuclei (Tanaka et al. 1995; Nandra et al. 1997).  The line width is
evidently rather small; we place an upper limit of $\sigma<0.16$
keV. Our data, unfortunately, do not have sufficient signal-to-noise
ratio to confirm the presence or absence of disklike kinematic
features in NGC 4579.  Future measurements of the line profile using
higher energy resolution and larger effective areas will be crucial to
investigate the origin of the Fe line and the structure of the
accretion disk in LLAGNs.

\acknowledgments

The authors are grateful to all the {\asca} team members. We also
thank an anonymous referee for useful comments. YT and KM thank Japan
Society for the Promotion of Science for support.

\figcaption[]{({\it a}) {\it ASCA} SIS (with circle) and GIS spectra
of NGC 4579.  The data sets from two observations in 1998 are
combined. The solid histograms are the best-fit model but the
normalization of the Gaussian component is set to be zero. The dotted
and dashed histograms represent the Raymond-Smith and the power-law
component, respectively. 
({\it b}) Ratio of the data to the best-fit model around Fe
lines in the 1998 observation and ({\it c}) 1995 observation. The
normalization of the Gaussian component is set to be zero. The crosses
with and without filled circle are SIS and GIS data, respectively.
The energy scale is not redshift corrected.  \label{fig-1} }

\figcaption[]{ 
Confidence contours ($\Delta \chi^2$=2.3, 4.6, and 9.1)
for the line energy and normalization. Dashed lines are for the 1995
observation and solid lines are for the combined spectrum of the two
1998 observations. The fitting model is Raymond-Smith + Power-law +
Gaussian. The energy scale is redshift corrected.
\label{fig-2}
}

\figcaption[]{
The equivalent width and intensity of the 6.4 keV and 6.7 keV line for
the 1995 and 1998 observations. A double Gaussian model with centroid
energies fixed at 6.4 keV and 6.7 keV is assumed for the Fe line.
\label{fig-3}
}

\footnotesize

%\small

\begin{table*}
\begin{center}
\begin{tabular}{ccccccccc}
\tableline \tableline
Data & \NH (galactic) & $kT$                 & Abundance     & \NH (hard component)
& $\Gamma$	&  Flux (2-10 keV)	& $\chi^2$/dof\\
        & [$10^{20}$~{\pcm}]    & [keV]         & [solar]       &
[$10^{20}$~{\pcm}]    &       & [$10^{-12}${\eps}{\pcm}]	&\\ 
\tableline
1995    & 3.1(f)      & $0.90^{+0.11}_{-0.05}$ & 0.50(f)        & $4.1\pm2.7$ & $1.72\pm0.05$   	& 4.3	& 192.4/201 \\
1998(1+2)& 3.1(f)      & 0.90(f) 		& 0.50(f)       & $3.8 (<13)$	& $1.81\pm0.06$	& 5.8	& 200.5/225 \\
1998(1)	& 3.1(f)      & 0.90(f) 		& 0.50(f)       & $14 (<28)$	& $1.86\pm0.09$	& 6.1	& 188.3/207 \\
1998(2)	& 3.1(f)      & 0.90(f) 		& 0.50(f)       & $3.4 (<13)$	& $1.84\pm0.08$& 5.3	& 201.1/200 \\
\tableline
\end{tabular}
\end{center}
\tablenum{1}
\caption{Results of spectral fitting to the SIS and GIS spectra of NGC 4579\label{tbl-1}}
\tablecomments{The fitting model is Raymond-Smith + Power-law + Gaussian. (f) 
in the table denotes a frozen parameter. The absorption column for the RS component is 
assumed to be the Galactic value. The quoted errors 
are at the 90\% confidence level for one interesting parameter.}
\end{table*}

%\clearpage

\begin{table*}
\begin{center}
\begin{tabular}{ccccc}
\tableline \tableline
model           & $E_L$         & $\sigma$      & EW 	& Intensity\\%$\chi^2$/dof\\
                & [keV]         & [keV]         & [eV]	& [10$^{-5}$photons s$^{-1}$ cm$^{-2}$]\\
\tableline
Single Gaussian  \\
%1995	& $6.73^{+0.13}_{-0.12}$ & $0.17^{+0.11}_{-0.12}$ & $490^{+180}_{-190}$ & \\ %192.4/201\\
%1998	& $6.39\pm0.09$	& $0.0 (<0.16)$	& $260\pm100$	& \\ %216.2/232\\
1995	& 6.73 (6.61--6.86)	& 0.17 (0.05--0.28)	& 490 (300-670)	 	& 1.90 (1.16--2.60)\\
1998(1+2)& 6.39 (6.30--6.48)	& 0.0 (0--0.16)		& 250 (155--355)	& 1.39 (0.86--1.97)\\
1998(1)& 6.37 (6.27--6.49)	& 0.0 (0--0.19)		& 270 (130--410)	& 1.54 (0.75--2.32)\\
1998(2)& 6.40 (6.24--6.62)	& 0.0 (0--0.28)		& 250 (110--445)	& 1.22 (0.55--2.27)\\
\tableline
Double Gaussian \\
1995	& 6.4(f)		& 0.0 (f)	& 0 (0--200)	 	& 0 (0--0.85)\\
	& 6.73 (f)		& 0.17 (f)	& 490 (300-670)		& 1.90 (1.16--2.60)\\
1998(1+2)& 6.4(f)		& 0.0 (f)	& 260 (155--360)	& 1.42 (0.85--1.97)\\
	& 6.73 (f)		& 0.17 (f)	& 0 (0--110)		& 0 (0--0.54)\\
\tableline
\end{tabular}
\end{center}
\tablenum{2}
\caption{Gaussian fits to the iron K line\label{tbl-2}}
\tablecomments{The line energies are at the source
rest frame. (f) in the table denotes a frozen parameter.
 The errors are 90\% confidence range for one interesting parameter.}
\end{table*}

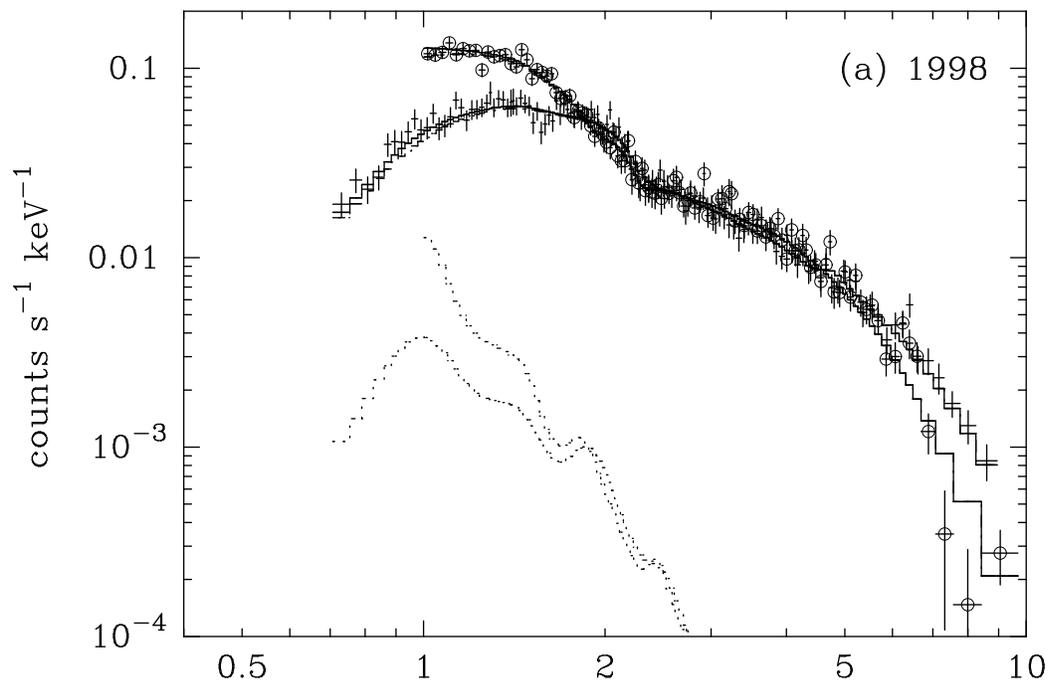
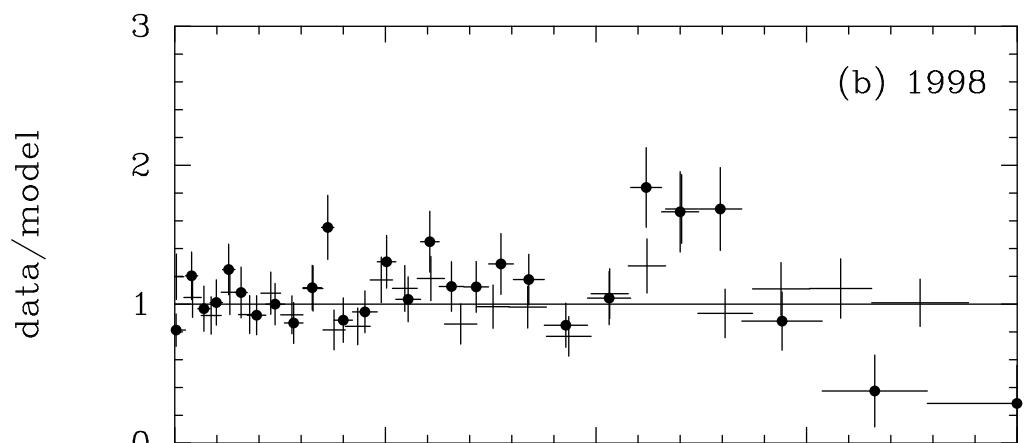
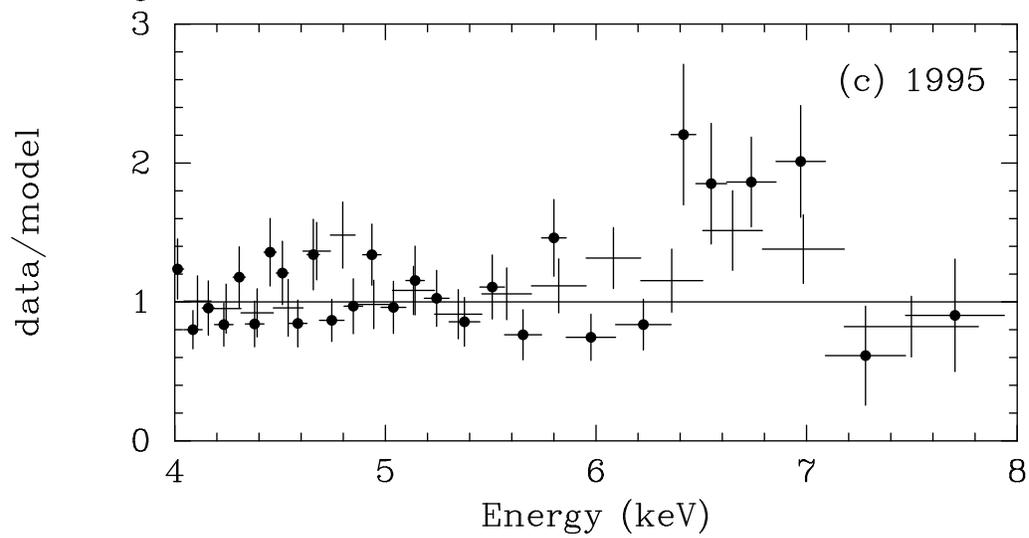
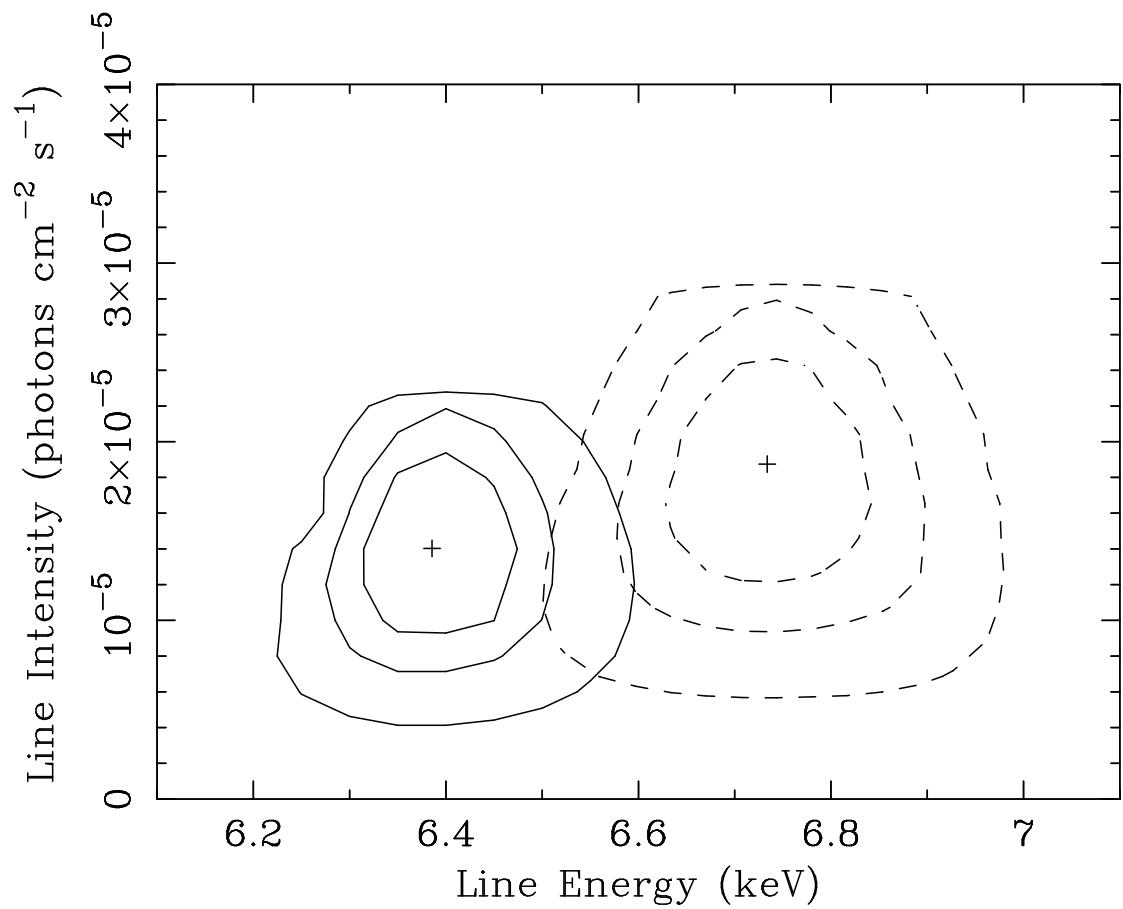
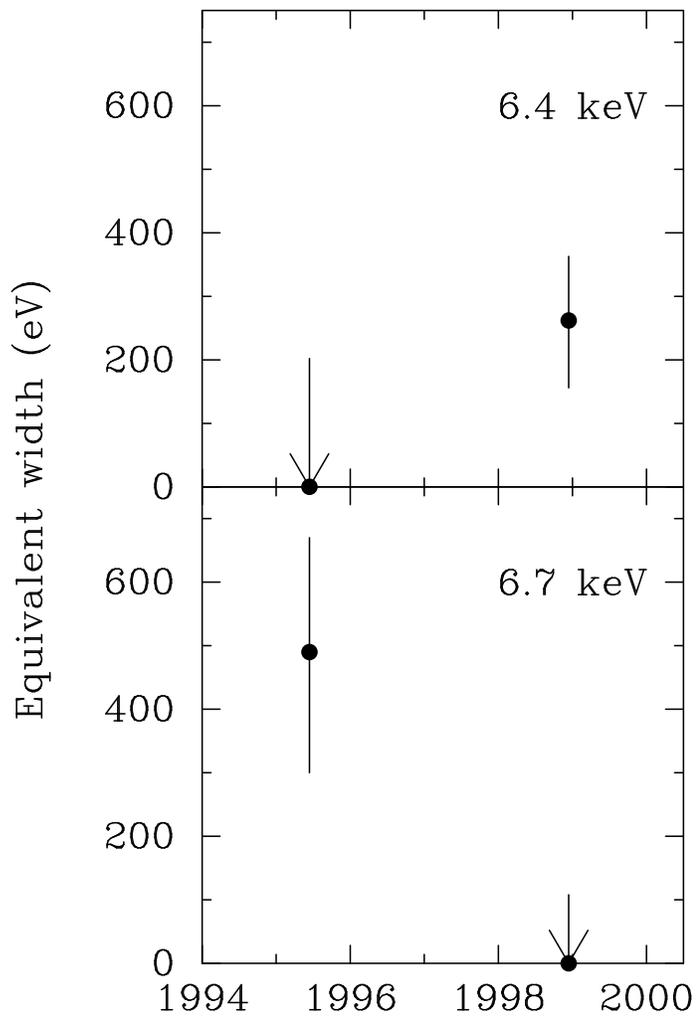
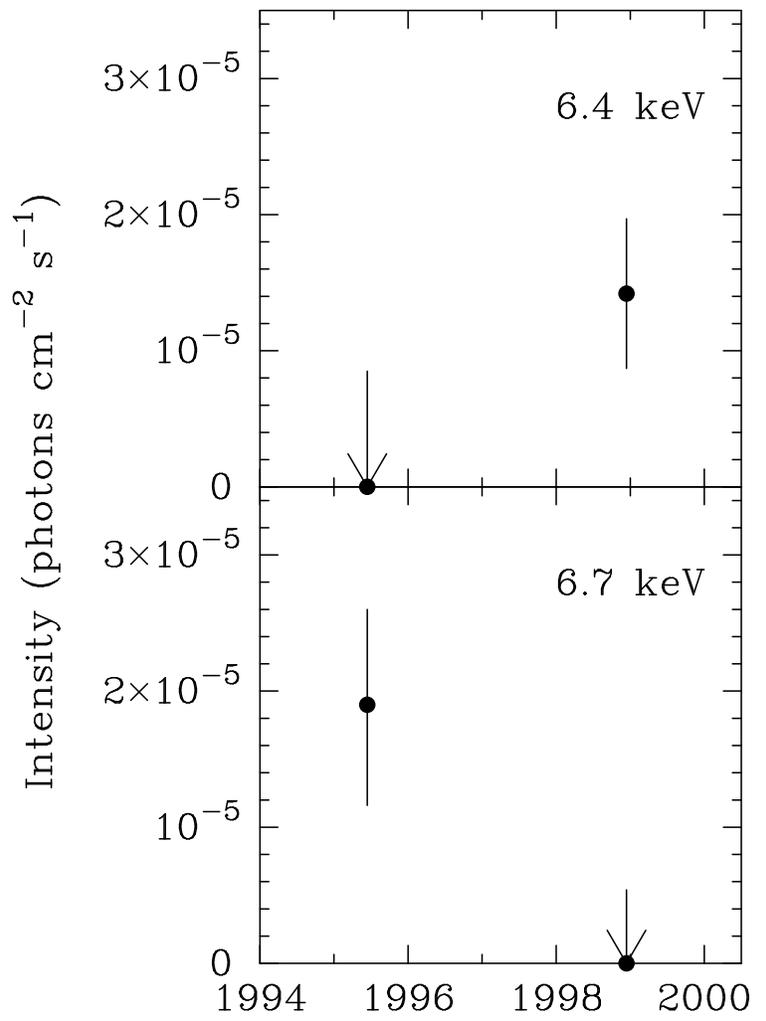
\end{document}